# A complete SPICE subcircuit-based model library for organic photodiodes

Anis Daami, Jérôme Vaillant, Romain Gwoziecki, Christophe Serbutoviez

*CEA/LITEN/DTNM/LCI, 17 Rue des Martyrs, Grenoble 38054 Cedex 9, France*

**Abstract:** A precise model library based on an equivalent circuit is presented and discussed. The model fully describes the light and bias behaviours of organic photodiodes realized on plastic substrates. The sub circuit modelling consists of a single SPICE LEVEL1 diode, a light power dependent series resistance, a shunt resistance. A combined light-power and voltage-controlled current source is also used to emulate the sensitivity behaviour of the photodiode. Moreover the model also permits designers to follow the technology deviations through the inclusion of worst-case corners. Finally a statistical model is included in order to allow designers run Monte Carlo simulations.

**Keywords:** *Spice modelling, Organic photodiode, Statistical model, Detectivity*

## 1. Introduction

Drastic efforts have been done in organic technologies to develop air stable materials and devices in the last decades. Beside the organic light-emitting diode (OLED) and the organic thin-film transistor (OTFT) developments, the organic photodiode (OPD) is one of the promising devices these last years [1–4]. The need of large area imagers for different applications [5–9] has become a great challenge to achieve. Unfortunately, organic device modelling has been mostly concentrated on OTFTs. Some SPICE models can even though be found in literature for inorganic and organic photo sensors [10–13]. In order to achieve a correct opto-electrical description of a pixel, accurate models of all its components have to be used. Furthermore these models have to be time-efficient in order to be easily integrated into SPICE simulators. In this work we focus on the modelling of the OPD in order to well describe both its electrical and optical behaviours. We present a DC/AC complete SPICE photodiode model library construction based on an equivalent circuit that allows fast and reliable simulations.

## 2. The model

### 2.1. The subcircuit

Fig. 1 shows the equivalent circuit used in our model. It consists of a SPICE LEVEL1 ideal diode (D), a series resistor (RS), a parallel resistor (RP) and a voltage dependent current source (G) which emulates the light behaviour of the photodiode. The subcircuit has three input pins. Two of them represent the anode and cathode of the photodiode. The third one called ''light'', when biased positively, is used to imitate the incident light power received by an OPD when illuminated. The bias applied to this later pin will then be the control voltage of the current source G. RS and RP resistors are also described by the SPICE LEVEL1 resistance model. This allows us later to model the temperature behaviour of the photodiode using the specific dedicated parameters in this later model.

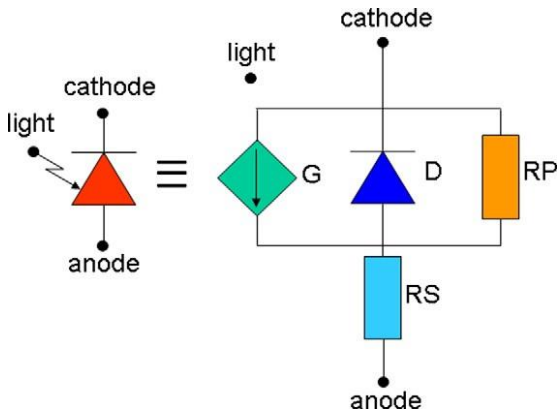

**Fig. 1.** Equivalent circuit of the photodiode.

### 2.2. DC modelling

#### 2.2.1. Dark behaviour

We show in Fig. 2 an OPD current–voltage (I–V) characteristic in dark conditions at a temperature of 25 C, with a surface of 0.0314 cm$^2$. The simulation using our model fits very well the photodiode measured dark current. In a first approximation we have supposed that the photodiode current is only area proportional and set to nil the D model periphery-dedicated parameters. Beside the RS value, no more than the saturation current (IS) and the emission coefficient (N) in the diode model have been tuned in order to adjust the ON current of the direct biased photodiode. It is evident that the internal series resistance parameter of the D model is put to nil as the ON current crowding is taken into account through the resistor RS. On the other hand RP value has been adjusted to fit the dark OFF current of the photodiode when it is reverse biased.

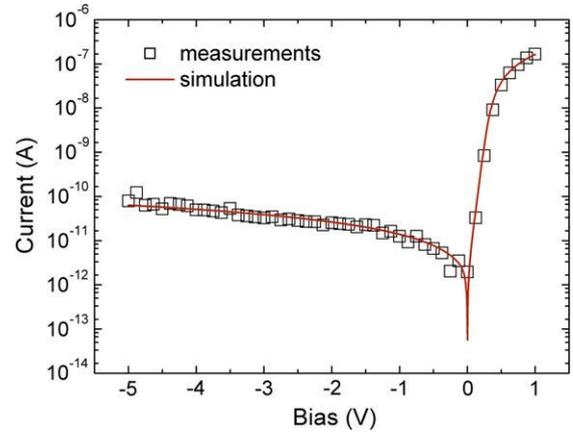

**Fig. 2.** Dark current–voltage characteristic of the organic photodiode: simulation versus measurements.

#### 2.2.2. Light behaviour

##### 2.2.2.1. Wavelength dependence.

External Quantum Efficiency (EQE) measurements have been realized on our OPD at wavelengths raging from 0.42 μm to 1.2 μm at a reverse bias of 1V. The OPD sensitivity (S) can then be deduced easily for each wavelength through the formula below:

$$S(\lambda) = \eta(\lambda)\frac{q\lambda}{hc} \qquad (1)$$

where $\eta(\lambda)$ stands for the EQE value at a wavelength value $\lambda$, $q$ the single electron charge and $h$ as the Planck's constant.

To represent the whole sensitivity spectrum of the OPD, we have chosen to divide the spectrum into three parts that have been empirically fitted each of them with a 6 polynomial that ensure a good description of the whole spectrum. We have also verified the good continuity of the values at the joining wavelength points. This method allows us to integrate easily the equations into the device model without having complicated calculus to run on a SPICE simulator.

We show in Fig. 3, the whole measured sensitivity and the simulated one at a reverse bias of 1 V, where we can see a quasi perfect superposition of both curves.

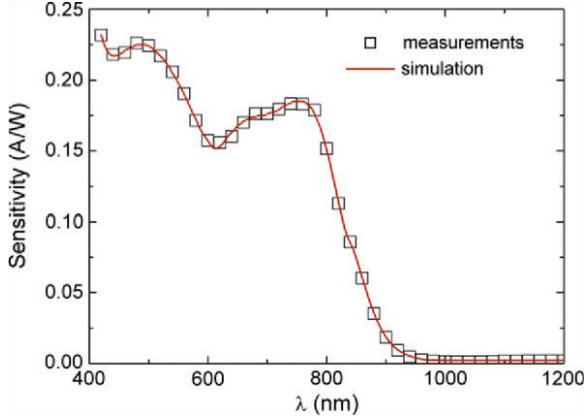

**Fig. 3.** Sensitivity spectrum of the organic photodiode: simulation versus measurements.

*2.2.2.2. Reverse bias dependence*

In Fig. 4 we have represented the calculated sensitivity of an OPD at three different wavelengths when reverse biased from 0 V to 5 V. One can notice that the sensitivity increases when going to higher absolute bias values. This effect can be of a great benefit for low sensitivity materials. We also observe that the increase of sensitivity seems independent of the chosen wavelength. Indeed we managed to adjust the three curves with a common simple equation:

$$S(\lambda, V) = S_\infty(\lambda) - B(\lambda)\exp(V) \quad (2)$$

where $S_\infty(\lambda)$ and $B(\lambda)$ are coefficients depending on the wavelength $\lambda$.

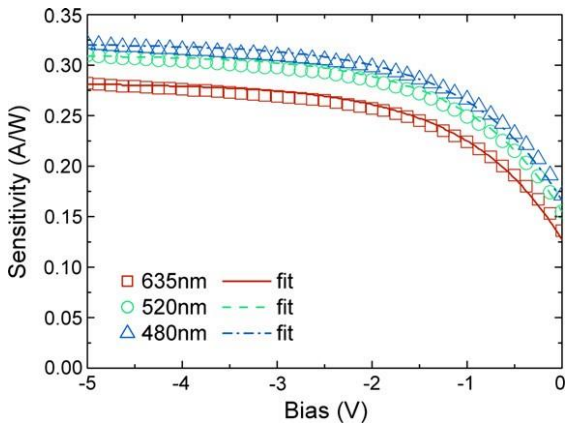

**Fig. 4.** Sensitivity bias dependence at three different wavelengths: 520 nm, 635 nm and 480 nm: measurements versus fitting equation.

We interpret $S_\infty(\lambda)$ as the maximum sensitivity that can be achieved at a certain wavelength for the OPD with a certain high reverse bias. As said above, we have observed that the sensitivity variation with the reverse bias seems to be independent of the wavelength which implies that the factor $B(\lambda)$ should not depend on $\lambda$. Nevertheless, we can easily think that, when the sensitivity of the OPD starts to diminish and falls to zero from a certain wavelength value (0.8 μm in our case), the bias dependence will also become negligible and will surely tend to zero. Therefore the coefficient B(k) has been modelled through Eq. (3) in order to take into account this probable effect.

$$B(\lambda) = \begin{cases} B_0 & \text{if } \lambda \leq \lambda_0 \\ B_0 \left(\frac{\lambda - \lambda_1}{\lambda_0 - \lambda_1}\right) & \text{if } \lambda_0 < \lambda \leq \lambda_1 \\ 0 & \text{if } \lambda > \lambda_1 \end{cases} \quad (3)$$

here, $\lambda_0$ is for the limit till which the reverse bias dependence is supposed to be wavelength independent and $\lambda_1$ the wavelength from which sensitivity quenches to zero.

We show in Fig. 5 an example of the modelled sensitivity reverse-bias dependence, by simulating the response of our OPD, on its whole light-response wavelength range, at two different reverse biases (1 V and 5 V).

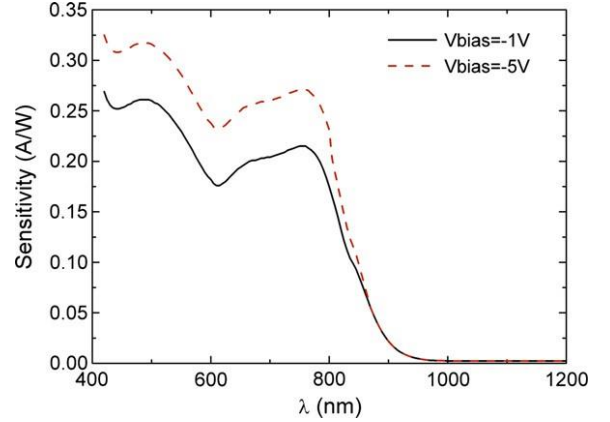

**Fig. 5.** Simulated sensitivity spectrum at two reverse biases.

*2.2.2.3. Series resistor light-dependence.*

We have observed an increase in the ON current of our fabricated OPD on plastic substrate when forward biased and illuminated simultaneously. To take into account this ON current increase under incident light we have implemented a light-dependent series resistor as shown below in the following equation:

$$RS = \begin{cases} RS_{dark} & \text{if } V(light) = 0 \\ RS_{light} & \text{if } V(light) > 0 \end{cases} \quad (4)$$

This implementation has no incidence on photodiode models that do require this double series resistor definition as it can be deactivated easily. Fig. 6 shows a comparison between measurements and simulation of I–V characteristics on an OPD having an area of 0.0314 cm² in dark and light conditions.

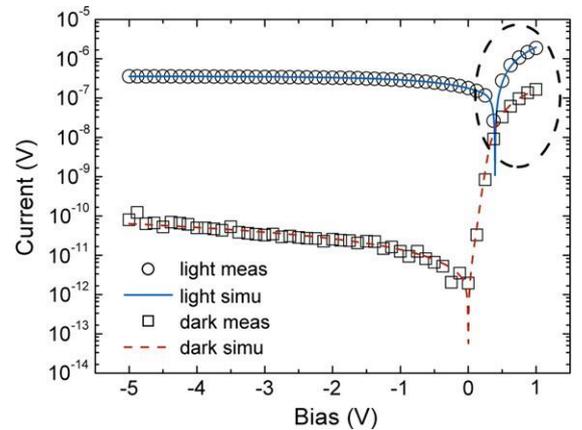

**Fig. 6.** Simulation versus measurements for dark and illuminated current–voltage characteristic of the photodiode.

## 2.3. AC modelling

### 2.3.1. Capacitance modelling

To have a complete description of the OPD electrical behaviour, capacitance–voltage (*C*–*V*) measurements have been realized at reverse biases ranging from 0.5 V to 4 V. By adjusting the capacitance related parameters of the diode model D we managed to reproduce by simulation the measured curve without many difficulties. The superposition of both curves is shown in Fig. 7.

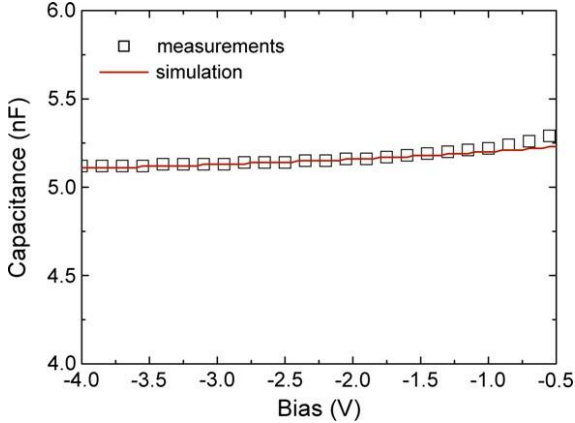

**Fig. 7.** Capacitance–voltage characteristic of the photodiode: simulation versus measurements

## 2.4. Temperature modelling

In order to adjust the temperature current behaviour of the OPD we have only used the RP and RS resistors temperature related parameters, respectively to reflect the dark OFF current and the ON current of the photodiode. Two temperature parameters of the ideal diode D have been also used to adjust its saturation current level. Fig. 8 shows *I*–*V* characteristics measured at two temperatures 25°C and 45°C on an organic photodiode having an area of 0.353 cm$^2$ and their corresponding SPICE simulations. We can easily observe that there is no discrepancy between the measured and simulated curves.

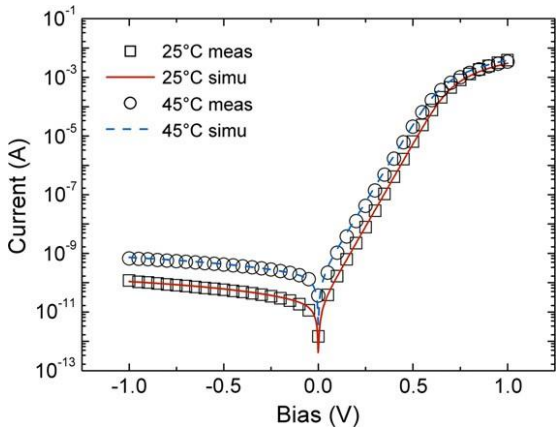

**Fig. 8.** Temperature dark current–voltage photodiode behaviour: simulation versus measurements.

## 3. The library construction

Based on the spreading of our OPD technology we have defined variations on identified model parameters of our subcircuit in order to allow to simulate the spread that can be observed on the output current of the photodiode. Two different modelling types can be used to reflect the spread of a technology: worstcase corners modelling and statistical modelling. We have chosen to use both types to construct a complete model library. This allows designers having the choice between simulating complex circuits rapidly in worstcase conditions or by longer Monte Carlo simulations.

### 3.1. Worstcase corners

We have defined five worstcase conditions taking into account dark and light conditions outputs. Each corner is defined by two letters; the first will define the ON/OFF currents and the reverse capacitance in dark conditions, the second will define the sensitivity level and thus the photodiode response in illuminated conditions. The typical corner TT has been chosen having the mean outputs and thus no variations are applied to this worstcase corner. Table 1 summarizes the output variations depending on which corner will be simulated. To clarify the definition of these corners we should mention that letters F, S and T stand for slow, fast and typical.

**Table 1**
Photodiode outputs variations depending on simulated worstcase corner.

| | SS | SF | TT | FS | FF |
|---|---|---|---|---|---|
| ON current | − | − | 0 | + | + |
| OFF current | + | + | 0 | − | − |
| Reverse capacitance | + | + | 0 | - | - |
| Sensitivity | - | + | 0 | - | + |

Fig. 9 respectively show simulated *I*–*V* characteristics in dark and illuminated conditions. Light conditions were fixed at $\lambda$ = 520 nm with an incident power of 36.8 µW/cm$^2$. As awaited from the definition of our corners, we observe that the FF corner gives the highest dynamic range between dark and light for a reverse biased photodiode.

Moreover, the reverse capacitance simulated for the three same corners and represented in Fig. 9 is giving the highest value for the SS corner which is coherent with our previous definitions.

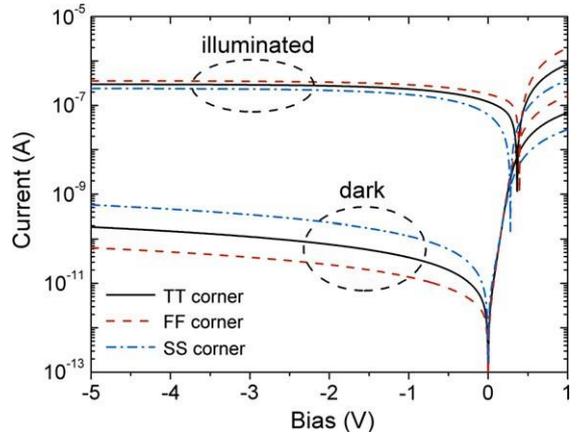

**Fig. 9.** Simulated current–voltage behaviour using worstcase corners in dark and illuminated conditions.

### 3.2. Statistical modelling

To be consistent with the worstcase corners, we have introduced Gaussian distributions on the same model parameters used for the worstcases construction. The standard deviations of the Gaussian distributions have been adjusted so that the $3\sigma$ output variations of the photodiode; ON/OFF currents, reverse capacitance and sensitivity; lay between the FF and the SS corners.

## 4. Noise equivalent power (NEP) and detectivity (D)

Using our OPD statistical model, we have quantified by simulation our photodiode noise equivalent power (NEP), which by definition is the minimum incident light power needed to generate a signal-to-noise ratio (S/N) equal to unity at a certain wavelength k. It is defined as:

$$NEP(\lambda) = \frac{I_N}{S(\lambda)} \quad (5)$$

Expressed in W/Hz$^{1/2}$, where $I_N$ is the noise current of the OPD expressed in A/Hz$^{1/2}$ and $S(\lambda)$ the photodiode sensitivity defined in Eq. (1).

For a reverse biased photodiode, the dominant noise is found to be the shot noise [3] defined as:

$$I_{shot} = \sqrt{2qI_{dark}B} \quad (6)$$

where q is the electron charge, $I_{dark}$ the dark current of the photodiode at the simulated reverse bias and B the bandwidth associated to a certain exposure time ET and approximately expressed as:

$$B \approx \frac{1}{2ET} \quad (7)$$

In our simulation, we have considered a bandwidth B = 1 Hz.

After a, 10,000 runs, Monte Carlo simulation, we show in Fig. 10, the statistical distribution of a photodiode NEP at a wavelength $\lambda$ = 520 nm and having a surface of 0.0314 cm$^2$. The mean NEP value extracted is $3.4 \times 10^{14}$ W/Hz$^{1/2}$ with a standard deviation $\sigma_{NEP} = 6.2 \times 10^{15}$ W/Hz$^{1/2}$.

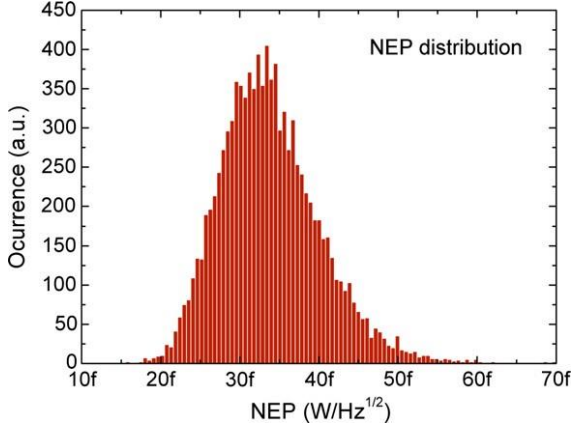

**Fig. 10.** Simulated noise equivalent power of the photodiode using the statistical model.

For a matter of comparison with other photodetectors, the detectivity parameter $D^*$ is defined as:

$$D^*(\lambda) = \frac{\sqrt{A}}{NEP(\lambda)} \quad (8)$$

Expressed in cm Hz$^{1/2}$ W$^{-1}$, where A is the photodiode area expressed in cm$^2$.

The same simulation shows that our photodiode has a mean detectivity $D^* = 6.1 \times 10^{12}$ cm.Hz$^{\frac{1}{2}}$.W$^{-1}$ with a standard deviation $\sigma_{D^*} = 1 \times 10^{12}$ cm.Hz$^{\frac{1}{2}}$.W$^{-1}$.

We also ran the same Monte Carlo simulation for wavelengths of 468 nm and 820 nm in order to better characterize the detectivity or our organic photodiode. To compare our results with literature data [3], we also simulated the detectivity distribution at a reverse bias of 1 V for the wavelength 468 nm.

We summarize in Table 2 the results of all evaluated NEP and $D^*$ concerning our organic photodiode and compare them to available literature data and to an inorganic silicon HAMAMATSU photodiode [14]. We can notice that our photodiode shows a quite high detectivity level even though to the inorganic silicon one is not yet achieved.

## 5. Conclusion

An organic photodiode complete SPICE model library has been demonstrated. A subcircuit description of the photodiode permitted the accurate modelling of its dark and light behaviour over its complete bias interval and wavelength sensitivity range. Based on the typical model, worstcase corners and statistical model have been built to allow a complete spread study if needed by designers. Through a Monte Carlo simulation of the detectivity based on our organic photodiode library we have shown that our OPD technology can provide quiet high performance photodetectors as compared to inorganic technologies that still remain a step upward.

**Table 2**
Summary and comparison table of statistical simulation results for noise equivalent power and detectivity.

| Reference | λ (nm) | Bias (V) | NEP/σNEP (W/Hz$^{1/2}$) | D$^*$/σD$^*$ (cm.Hz$^{1/2}$.W$^{-1}$) |
|---|---|---|---|---|
| This work | 820 | 5 | 5.1 10$^{14}$/9.3 10$^{15}$ | 3.6 10$^{12}$/6.4 10$^{11}$ |
|  | 520 | 5 | 3.4 10$^{14}$/6.2 10$^{15}$ | 6.1 10$^{12}$/1.0 10$^{12}$ |
|  | 468 | 5 | 2.9 10$^{14}$/5.4 10$^{15}$ | 6.2 10$^{12}$/1.1 10$^{12}$ |
|  | 468 | 1 | 1.7 10$^{14}$/3.2 10$^{15}$ | 11 10$^{12}$/2.0 10$^{12}$ |
| [3] | 468 | 1 | 2.8 10$^{14}$/– | 7.0 10$^{12}$/– |
| [14] | 550 | 1 | 0.67 10$^{14}$/– | 39 10$^{12}$/– |